\documentclass[twocolumn,showpacs,preprintnumbers,amsmath,amssymb,prb,aps,nofootinbib]{revtex4-1}
\usepackage{graphicx}
\usepackage{dcolumn}
\usepackage{bm}
\begin{document}

\title{Raman studies on a heavily distorted polycarbonate sample $-$ Raman-Untersuchungen an einer stark deformierten Polycarbonat-Probe}

\author{J. Debus} \email[Corresponding author: ]{joerg.debus@tu-dortmund.de}
\affiliation{Experimentelle Physik 2, Technische Universit\"at Dortmund, 44227 Dortmund, Germany}
\author{D. Dunker}
\affiliation{Experimentelle Physik 2, Technische Universit\"at Dortmund, 44227 Dortmund, Germany}

\begin{abstract}
Differently distorted areas on a polycarbonate sample are studied by means of Raman spectroscopy. The various Raman lines, whose energy shifts range between 200 and 3200 cm$^{-1}$, are compared for different sample positions with respect to their spectral position, intensity and line width. While a double bonding does not indicate a change in the structural characteristics of the distorted polycarbonate, CH-stretching modes show strong Raman intensity differences induced by mechanical stress. Experimental setup modifications are outlooked.

Eine Polycarbonat-Probe wird an unterschiedlich stark deformierten Stellen mittels Raman-Spektroskopie untersucht. Die verschiedenen Raman-Linien mit Raman-Verschiebungen in einem Bereich von 200 bis 3200 cm$^{-1}$ werden f\"ur die unterschiedlichen Probenstellen hinsichtlich ihrer spektralen Position, Intensit\"at und Breite verglichen. Aus den experimentellen Resultaten geht hervor, dass Doppelbindungen nur unzureichend als Indikator f\"ur strukturelle Ver\"anderungen innerhalb des deformierten Polycarbonats dienen. Hingegen zeigen CH-Streckungen deutliche Raman-Intensit\"atsunterschiede in Abh\"angigkeit von dem Polycarbonat-Verspannungsgrad. Ein Ausblick auf experimentelle Erweiterungen wird gegeben.
\end{abstract}

\maketitle

Die Raman-Spektroskopie erm\"oglicht die Untersuchung von chemischen Zusammensetzungen und molekularen Strukturen von Polymeren auf einer mikroskopischen Skala; mikroskopische Eigenschaften k\"onnen makroskopische Parameter, wie beispiels\-weise die materialspezifische Dehnungsst\"arke, beeinflussen. Konformationelle Defekte, die Ver\"anderungen von Molek\"ulanordnungen infolge externer Einfl\"usse entsprechen und in Polymeren in der Regel auftreten, werden in Form von geordneten und ungeordneten Molek\"ulketten beschrieben. Diese wiederum werden durch lokale Parameter, zum Beispiel durch die re\-lative Besetzung von isomerischen Zust\"anden entlang einer zentralen Bindungsrichtung, definiert\cite{weber}. Solche konformationellen Defekte spiegeln sich in Frequenz\-verschiebungen und gegebenenfalls in Verbreiterungen von Raman-Signalen wider; die Absch\"atzung von lokalen Parametern ist m\"oglich. Ebenso zeigen sich die Auswirkungen von Temperatur\"anderungen auf polymere Strukturen in Raman-Spektren, bedingt durch \"Anderungen in der konformationellen Verteilung entlang von molekularen Ketten. Typische Raman-Signaturen werden Gruppen von Ketten-Konformationen zugeordnet. Ihre \"Anderungen dienen zur Charakterisierung polymerer Strukturen. W\"ahrend Beugungsmethoden Informationen \"uber langreichweitig geordnete Systeme liefern, adressiert die Raman-Spektroskopie die lokale Ordnung bzw. Unordnung. Au\ss erdem h\"angt die Raman-Intensit\"at von \"Anderungen in der Polarisierbarkeit polymerer Substrukturen ab; sie ist verh\"altnism\"a\ss ig gro\ss\ f\"ur Kohlenstoffbindungen, sofern sie die polymere Basis bilden. Polarisierte Raman-Streuung kann zahlreiche Informationen \"uber die Orientierung von Kettensegmenten und die Zuordnung von bestimmten B\"andern liefern. Wenn ein Polymer im geschmolzenen Zustand extrudiert wird, treten Deformationen \"uberwiegend w\"ahrend des \"Ubergangs in die feste Phase auf. Die Art und Weise der mikro- und makroskopischen Orientierungen h\"angt von dem Relaxationsprozess und der Spannung w\"ahrend des Erstarrens ab. Typischerweise bildet sich ein geringer Orientierungsgrad aus, abh\"angig von dem Temperaturprofil in unterschiedlichen Bereichen der polymeren Probe. Die Raman-Spektroskopie liefert eine Charakterisierung von den Orientierungen polymerer Segmente.

\begin{figure*}[t]
\centering
\includegraphics[width=12.5cm]{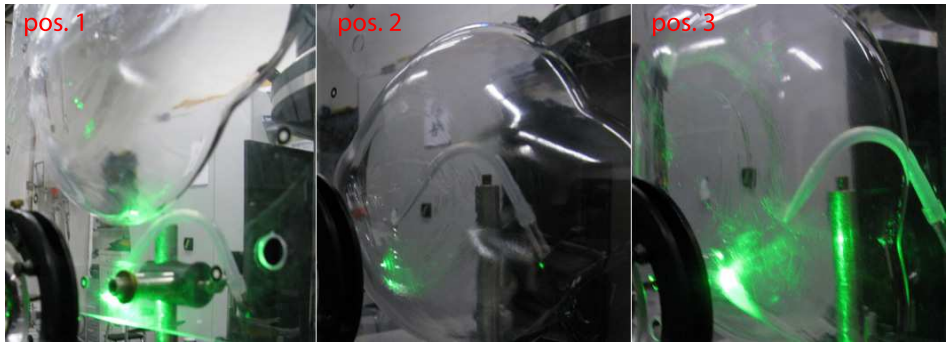}
\caption{Fotoaufnahmen der illuminierten Stellen der Polycarbonat-Probe. Die drei unterschiedlich verspannten Bereiche werden mit Position 1, 2 und 3 bezeichnet. Die Position 1 liegt in einem unbearbeiteten (spannungsfreien) Bereich, gem\"a\ss\ des \"au\ss eren Erscheinungsbildes weist die Position 2 einen mittleren und Position 3 einen starken Verspannungsgrad auf.}
\end{figure*}

Im Folgenden werden der experimentelle Aufbau, eine Analyse der Raman-Messdaten und abschlie\ss end ein Ausblick hinsichtlich experimenteller Optimierungsm\"oglichkeiten dargelegt. Das Untersuchungsobjekt der Raman-Messungen war eine stark deformierte Probe aus Polycarbonat.

\subsection{Experimentelle Details}
Die Raman-Messungen wurden mit einem gepulsten Nd:YVO-Laser, betrieben in der zweiten Harmo\-nischen ($\lambda_{\text{exc}}=532\,$nm), durchgef\"uhrt. Die Raman-Signale wurden mit einem dreistufigen Monochromator und einer stickstoffgek\"uhlten CCD-Kamera aufgenommen. Die Polycarbonat-Probe wurde in der R\"uckstreuungsgeometrie ohne Polarisationsoptik untersucht. Der Tripel-Monochromator arbeitete in dem subtraktiven Modus, in dem die ersten beiden Stufen als Laserfilter dienten und die Dispersion ausschlie\ss lich von der dritten/letzten Stufe \"ubernommen wurde. Die maximale spektrale Aufl\"osung dieser Konfiguration lag unterhalb von $1\,$cm$^{-1}$. Ein st\"orender Einfluss von Laserstreulicht konnte nicht festgestellt werden, obwohl die Zwischenspalte verh\"altnism\"a\ss ig weit ge\"offnet (500~$\mu$m) waren, um die Intensit\"at der Raman-Signale zu erh\"ohen. Die Messungen erfolg\-ten bei Raumtemperatur. Die R\"uckseiten der illuminierten Probenstellen wurden mit kaltem Stickstoffgas gek\"uhlt, um eine thermische Besch\"adigung des Polycarbonats durch den Laserstrahl zu vermeiden. Die Laserleistung betrug etwa 26~mW.

\subsection{Analyse der Raman-Spektren}
Die Analyse der Raman-Spektren f\"ur drei illuminierte Stellen der Polycarbonat-Probe umfasst die Charakte\-risierung der Raman-Signale und den Vergleich der Merkmale der jeweiligen Signale, um Aussagen \"uber die Auswirkungen von Deformationen des Polycarbonats auf die Raman-Signale treffen zu k\"onnen.

Es wurden drei unterschiedliche Bereiche der Probe mit dem Laser bestrahlt, nachfolgend werden sie mit Position 1, 2 und 3 gekennzeichnet. Entsprechend dem \"au\ss eren Erscheinungsbild der untersuchten Berei\-che wird zwischen einem vernachl\"assigbaren (pos. 1), mittleren (pos. 2) und hohen (pos. 3) Verspannungsgrad unterschieden. Die laserbestrahlten Stellen sind aus Gr\"unden der \"Ubersichtlichkeit und Reproduzierbarkeit fotografiert worden, siehe Abbildung 1.

Die Charakterisierung der beobachteten Raman-Signale basiert auf Ergebnissen aus verschiedenen Ver\"offentlichungen \cite{kul, janz, dybal, lee}. Den Ausgangspunkt der Signal\-identifizierung stellt die molekulare Struktur von Polycarbonat dar. Polycarbonat setzt sich haupts\"achlich aus Kohlenstoff-Sauerstoff(C$-$O, C$=$O)-, Kohlenstoff-Wasserstoff(CH$_3$)- und auch Kohlenstoff-Kohlenstoff(C$-$C)-Bindungen zusammen. Die Schwingungsfrequenzen bei der Raman-Spektroskopie h\"angen vorwiegend von Bindungsstreckungen und Bindungskr\"ummungen ab, die durch entsprechende Kraftkonstanten beschrieben werden. Konformationelle \"Anderungen \"au\ss ern sich nur geringf\"ugig in \"Anderungen der Kraftkonstanten. Daher \"andern sich die Schwingungsfrequenzen nur um eine oder zwei Wellenzahlen\cite{lee}, solche Differenzen k\"onnen mit dem verwendeten Raman-Setup detektiert werden.

Die aufgenommenen Raman-Spektren f\"ur die drei untersuchten Probenstellen sind in der Abbildung 2 dargelegt. Die Spektren weisen dieselben Raman-B\"ander auf, allerdings unterscheiden sie sich in ihrer Intensit\"at, Breite und teilweise auch in ihrer Frequenz. Diese Merkmale werden mittels Gau\ss -Anpassungen der Raman-Signale ermittelt, wobei die Genauigkeit auf 1~cm$^{-1}$ festgelegt/ reduziert wird. Von der hohen Genauigkeit (0,1~cm$^{-1}$) einer differentiellen Raman-Spektroskopiemethode wird in dieser ersten Untersuchungsphase abgesehen. Insgesamt sind 20 hinreichend ausgepr\"agte Raman-Signale feststellbar. In Tabelle 1 sind ihre Merkmale und ihre m\"oglichen physikalischen Urspr\"unge aufgef\"uhrt.

Die Raman-Spektren weisen ein Charakteristikum hinsichtlich der Abh\"angigkeit der relativen Signalintensit\"at von dem Verspannungsgrad auf: Je h\"oher der $-$ver\-mutete$-$ Verspannungsgrad ist, desto intensiver ist das Raman-Signal. Die ermittelten relativen Intensit\"aten ber\"ucksichtigen den gr\"o\ss eren Fluoreszenz-Hintergrund infolge des st\"arkeren Streulichts bei verspannten Stellen. Ob sich die Frequenz bei ver\"andertem Verspannungsgrad systematisch \"andert, l\"asst sich im Rahmen dieser ersten Messreihe nicht manifestieren. Dennoch legen die beobachteten Frequenz\"anderungen die Vermutung nahe, dass sich Deformationen in der Polycarbonat-Probe durch eine Frequenz\"anderung bestimmter Raman-Signale \"au\ss ern. Welche Schwingungsb\"ander hierf\"ur besonders geeignet sind, l\"asst sich bisher nur sch\"atzen: Die Frequenzen der C-O-Ringstreckungen reagieren scheinbar sehr schwach auf externe Einfl\"usse, w\"ahrend der Einfluss auf die CH-Bindungen st\"arker zu seien scheint. Einen weiteren Indikator zur Feststellung von strukturellen \"Anderungen stellt die Halbwertsbrei\-te (HWB) der Raman-Signale dar. Relativ brei\-te Peaks sind in dem spektralen Bereich zwischen 1200 und 1600~cm$^{-1}$ zu erkennen. Sie k\"onnen C-O-Streckschwingungen sowie Phenyl-Ring-Atmungsmoden (CH$_3$-Deformationen) zugeordnet werden. Ihre Breite weist auf das Vorhandensein von mehreren polymeren Konformationen hin. Diese Schwingungen k\"onnen eine wichtige Rolle hinsichtlich der Identifizierung morphologischer \"Anderungen, bedingt durch die Verformungen des Polycarbonats, einnehmen (vgl. Aging-Mechanismus \cite{lee}). 

\begin{figure*}[t]
\centering
\includegraphics[width=12.5cm]{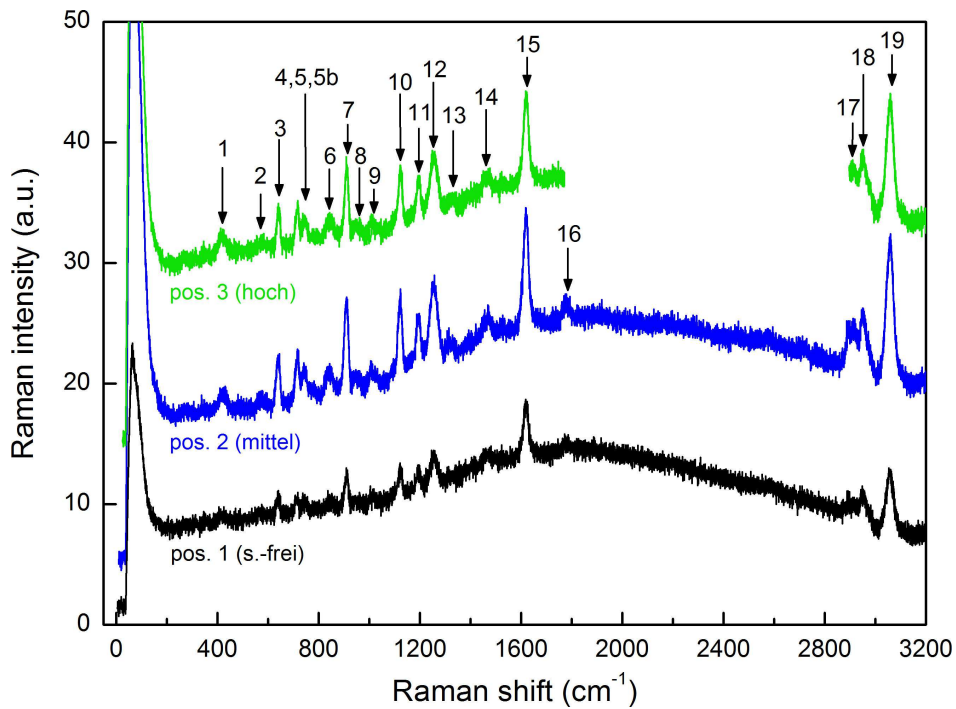}
\caption{Raman-Spektren von den illuminierten Stellen (Pos. 1, 2 und 3) der Probe aus Polycarbonat. Die Raman-Signale sind nummeriert, ihre Charakterisierung findet sich in Tabelle 1. F\"ur die Pos. 3 (gr\"un) wurden nur die linienreichen Frequenzbereiche aufgenommen.}
\end{figure*}

Ein Merkmal der Carbonyl-Gruppe ist die Streckung der Kohlenstoff-Sauerstoff-Doppelbindung(C=O) aufgrund der starken \"Anderung des Polarisierbarkeitstensors. Die typische Wellenzahl der korrespondierenden Schwingung liegt zwischen 1700 und 1780~cm$^{-1}$. Die unterschiedlichen Stellen der Polycarbonat-Probe zeigen ein schwaches Raman-Signal in diesem Frequenzbereich (Peak 16 bei etwa 1779~cm$^{-1}$). Die Intensit\"atsschw\"ache des Carbonyl-Signals kann auf einem geringen Streuquerschnitt der C=O-Bindung beruhen. Die geringe Raman-Aktivit\"at der Schwingungsmode ist vermutlich auf eine starke St\"orung der C=O-Bindungen zur\"uckzuf\"uhren; Spannungen f\"uhren zu Deformationen der Kohlen-/Sauerstoff-Doppelbindungen, wodurch die inelastische Streuwahrscheinlichkeit mit einfallenden Photonen abnimmt. Zudem nimmt die Frequenz der C=O-Streckschwingung um etwa 1~cm$^{-1}$ bei mittlerem Verspannungsgrad ab. Durch einen Deformierungsprozess k\"onnen starke Kohlenstoff-Sauerstoff-Bindungsdehnungen stattfinden, so dass sich der Abstand zwischen Kohlen- und Sauerstoff vergr\"o\ss ert und sich die Raman-Frequenz entsprechend verringert. Allerdings zeigt die untersuchte spannungsfreie Stelle ebenfalls ein schwaches Raman-Signal. Daher ist es relativ unwahrscheinlich, den Grund der Intensit\"atsschw\"ache in einem starken Verspannungsgrad zu suchen. Viel\-leicht dienen die Doppelbindungen nur unzureichend als Indikator f\"ur strukturelle Ver\"anderungen innerhalb des deformierten Polycarbonats?

Hingegen ist der relative Intensit\"atsunterschied zwischen der asymmetrischen (2950~cm$^{-1}$) und der symmetrischen (2900~cm$^{-1}$) CH-Streckung ein sensitiver Indikator f\"ur strukturelle \"Anderungen. Beispielsweise k\"onnen die Signale in dem Bereich der Methylen-Streckschwingungen um 2900~cm$^{-1}$ auf Fermi-Resonanzen zwischen symmetrischen Anregungen von fundamentalen Streckschwingungen mit niedriger und hoher Frequenz zur\"uckgef\"uhrt werden\cite{weber}. Fermi-Resonanzen treten auf, wenn zwei unterschiedliche Schwingungen von derselben Symmetrieart nahezu identische Energien besitzen. Die Sensitivit\"at der CH-Streckschwingungen gegen\"uber Struktur\"anderungen resultiert aus einem inh\"arenten Verst\"arkungsfaktor. Niederenergetische Schwingungen sind f\"ur ihre Sensitivit\"at gegen\"uber Ketten-Konformationen und Unterschiede in der Kettendichte bekannt. Infolge der Fermi-Resonanz k\"onnen geringe Frequenz\"anderungen in der niederfrequenten Mode signifikante \"Anderungen in dem hochfrequenten Spektrum bewirken.

\subsection{Fazit und Ausblick}
Zum einen zeigt sich f\"ur die drei unterschiedlichen Stellen der Polycarbonat-Probe eine Intensit\"atszunahme bei einer Vergr\"o\ss erung des Verspannungsgrades f\"ur das jeweilige Raman-Signal (Peak 17 und 18). Aus den Intensit\"atsverh\"altnissen (Intensit\"at von Peak 17 relativ zu der von Peak 18) geht hervor, dass mit erh\"ohtem Verspannungsgrad die Intensit\"at des Raman-Signals der symmetrischen CH-Streckschwingung etwas st\"arker zunimmt. Ein \"ahnliches Verhalten kann bei polymeren Strukturen gefunden werden, wenn sie einer erh\"ohten Temperatur ausgesetzt werden. Ein weiterer Unterschied zwischen den beiden Raman-Signalen deutet sich in ihren Frequenzen an, die stark variieren. Eine systematische Frequenz\"anderung l\"asst sich allerdings aus der geringen Anzahl an untersuchten Stellen nicht aufzeigen.

Die Detektion von den typischen Raman-Signalen der Carbonyl- und Methylen-Gruppe in deformiertem Polycarbonat unter Anregung mit moderaten Laserleistungen ist mit dem gew\"ahlten experimentellen Aufbau ohne gro\ss en Aufwand m\"oglich. Die Charakterisierung der einzelnen Raman-Signale funktioniert ebenfalls, Deformationsauswirkungen auf die innere Struktur der Polycarbonat-Probe k\"onnen festgestellt werden. Die Analyse der Raman-Spektren als auch die Lite\-raturrecherche zeigten, dass f\"ur eine definitive Be\-stimmung von konformationellen \"Anderungen einige experimentelle Aspekte optimiert werden sollten.

Der Einfluss von plastischen Deformationen auf die innere Struktur von Polymeren, speziell von amorphem Polycarbonat, kann mittels der Raman-Spektroskopie im niederfrequenten Bereich (2 $-$ 200~cm$^{-1}$) ermittelt werden\cite{mermet}. Die breiten, niederfrequenten Raman-B\"ander enthalten Informationen \"uber die Orientierung von makromolekularen Verbindungen und \"uber das anomale Verhalten der spezifischen W\"arme, hier\"uber k\"onnen auf m\"ogliche strukturver\"andernde Prozesse w\"ahrend der Deformation geschlossen werden. Mittels des niedrigen Frequenzbereichs k\"onnen auch zwischenmolekulare Wechselwirkungen untersucht werden, wodurch Informa\-tionen \"uber das freie Volumen innerhalb des Polycarbo\-nats gewonnen werden k\"onnen. Da die Raman-Signale unterhalb von 200~cm$^{-1}$ wesentlich sensitiver gegen\"uber Deformationen sind, sollte dieser Bereich in zuk\"unftigen Messungen ebenfalls analysiert werden. Dies setzt eine st\"arkere Unterdr\"uckung des Laserstreulichts und damit kleinere Zwischenspalte voraus. Die damit einhergehende Reduzierung der Signalintensit\"aten kann durch eine h\"ohere Aufnahmezeit der Spektren kompensiert werden. Eine Erh\"ohung der Laserleistung ist nur in gewissem Ma\ss e m\"oglich, da eine Besch\"adigung und zus\"atzliche Beeinflussung der illuminierten Probenstelle zu vermeiden ist. 

Au\ss erdem sollte die Idee der differentiellen Raman-Spektroskopie in Betracht gezogen werden (vgl. Lee et al.\cite{lee}). Hierbei wird das unbehandelte Probenst\"uck in einer Messreihe (nahezu simultan ohne Ver\"anderung der spektralen Position des Monochromators) mit der verspannten Probe untersucht. Nach Bildung der Differenz aus beiden Raman-Spektren k\"onnen Abweichungen direkt verdeutlicht werden. Alternativ zu der Subtraktion k\"onnen die Raman-Signale parametrisch angepasst und anschlie\ss end die Frequenzen und relativen Intensit\"aten miteinander verglichen werden. Die spektrale Genauigkeit bewegt sich hierbei in dem Be\-reich von 0,1~cm$^{-1}$. Da sich ther\-mische Einfl\"usse auf Polycarbonat-Strukturen in geringen Frequenzverschiebungen ($\approx$ 1~cm$^{-1}$) zeigen, ist diese Methode zu empfehlen. 

Eine weitere Option ist die optische Anregung in dem ultravioletten Spektralbereich ($\approx$~355~nm). Da die Adressierung von kohlenstoffartigen sp$^3$-Hybridorbitalen mit hochenergetischem Licht wahrscheinlicher ist, k\"onnen zus\"atzliche Informationen \"uber Kohlenstoff-Kohlenstoff-Wechselwirkungen gewonnen werden. Ver\"offentlichungen \"uber diese hochenergetische Anregung von Polymeren konnten wir bisher nicht finden. Hinsichtlich des Anregungsmechanismus k\"onnten auch Reflexionsmessungen im nahinfraroten Spektralbereich durchgef\"uhrt werden. Diese Methode ist neben der Raman-Spektroskopie ebenfalls zur Charakterisierung von polymeren Strukturen etabliert\cite{weber}. 

\begin{figure*}[t]
\centering
\includegraphics[width=13.1cm]{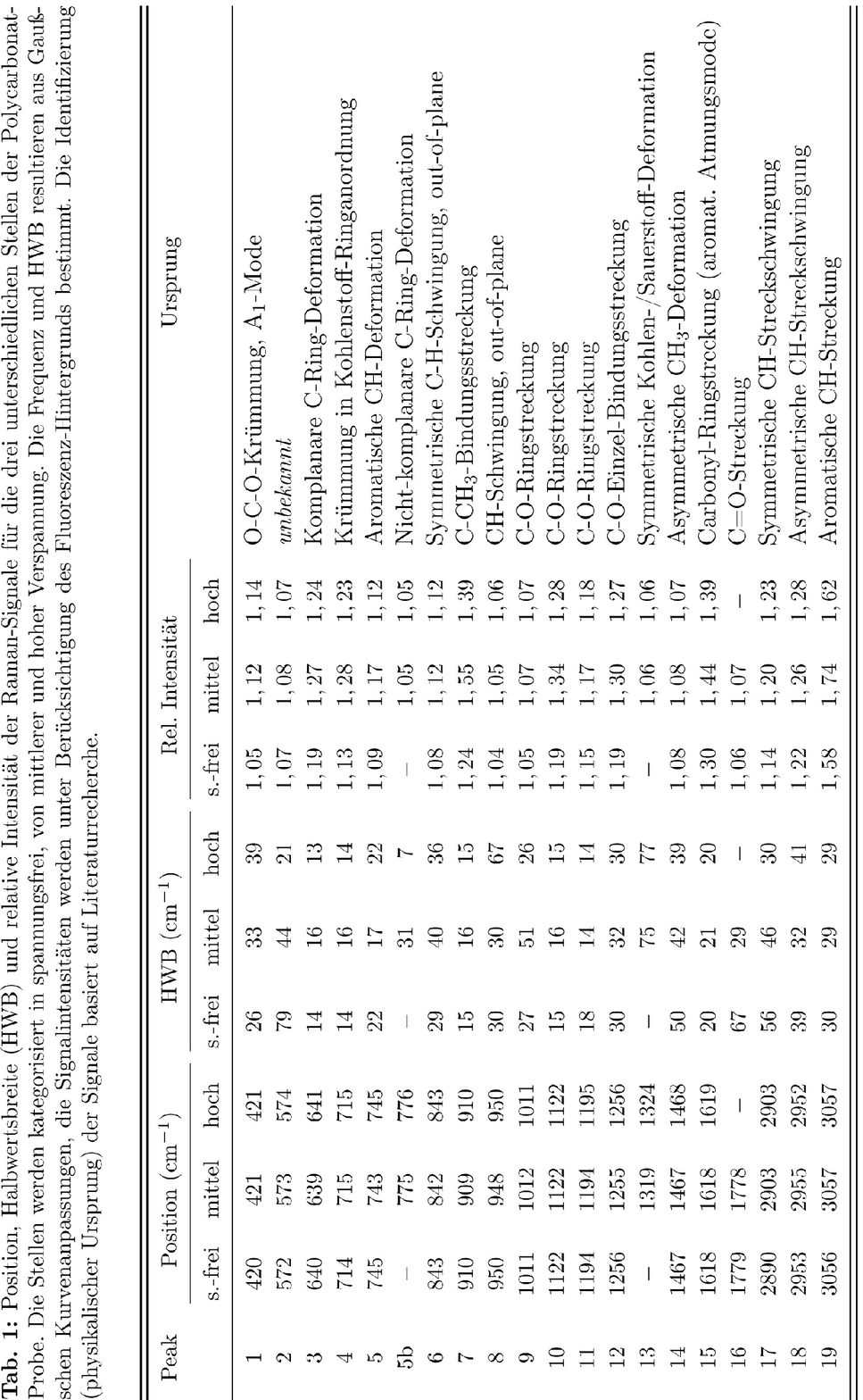}
\end{figure*}

Aufgrund der Sensitivit\"at der Raman-B\"ander auf die Polarisierbarkeit molekularer Strukturen, der Abh\"angigkeit der Frequenz der Raman-aktiven Schwingungsmoden von der Orientierung konformationeller Ketten, k\"onnen polarisationsabh\"angige Raman-Messungen Informationen \"uber strukturelle Ver\"anderungen liefern. Des Weiteren ist die Verwendung eines einzelnen Monochromators ohne starke Streulicht\-unterdr\"uckung zu \"uberpr\"ufen und die N\"utzlichkeit einer solchen Anordnung f\"ur in-situ Materialcharakterisie\-rungen einzusch\"atzen. Hierf\"ur biete sich die Sensitivit\"at der CH-Streckschwingungen bei 2900~cm$^{-1}$ gegen\"uber strukturellen \"Anderungen an, diese sollten mit einem einfachen Monochromator problemlos detektiert werden k\"onnen.

\end{document}